\documentclass[journal]{IEEEtranTIE}
\usepackage{graphicx}
\usepackage{cite}
\usepackage{picinpar}
\usepackage{amsmath}
\usepackage{url}
\usepackage{flushend}
\usepackage[latin1]{inputenc}
\usepackage{colortbl}
\usepackage{soul}
\usepackage{multirow}
\usepackage{pifont}
\usepackage{color}
\usepackage{alltt}
\usepackage[hidelinks]{hyperref}
\usepackage{enumerate}
\usepackage{siunitx}
\usepackage{breakurl}
\usepackage{epstopdf}
\usepackage{pbox}

\begin{document}
\title{	Qi Standard Metasurface for Free-Positioning and Multi-Device Supportive Wireless Power Transfer}

\author{
	\vskip 1em
	
	Hanwei Wang, \emph{Member, IEEE},
	Joshua Yu,
    Xiaodong Ye, \emph{Student Member, IEEE},
    Yun-Sheng Chen,\emph{Member, IEEE},
	\\ and Yang Zhao, \emph{Member, IEEE}

	\thanks{
	
		Manuscript received February 3, 2023; revised Month xx, xxxx; accepted Month x, xxxx.
		This work was partially supported by Jump ARCHES endowment through the Health Care Engineering System Center, Dynamic Research Enterprise for Multidisciplinary Engineering Sciences (DREMES) at Zhejiang University and University of Illinois Urbana-Champaign, and NIGMS R21GM139022. (Corresponding author: Yang Zhao, email: yzhaoui@illinois.edu)
  
		Hanwei Wang, Xiaodong Ye, and Yang Zhao are with Nick Holonyak Micro and Nanotechnology Laboratory and the Electrical and Computer Engineering Department, University of Illinois at Urbana-Champaign, Urbana, 61801, US. Joshua Yu is with the Electrical and Computer Engineering Department, University of Illinois at Urbana-Champaign, Urbana, 61801, US. Yun-Sheng Chen is with the Beckman Institute for Advanced Science and Technology and the Electrical and Computer Engineering Department, University of Illinois at Urbana-Champaign, Urbana, 61801, US. 
	}
}

\maketitle
	
\begin{abstract}
Free-positioning and multi-user supportive wireless power transfer systems represent the next-generation technology for wireless charging under the Qi standard. Traditional approaches employ multiple transmitting coils and multi-channel driving circuits with active control algorithms to achieve these goals. However, these traditional approaches are significantly limited by cost, weight, and heating due to their relatively low efficiency. Here, we demonstrate an innovative approach by using a metasurface to achieve free-positioning and multi-user compatibility. The metasurface works as a passive device to reform the magnetic field and enables high-efficiency free-positioning wireless power transfer with only a single transmitting coil. It shows up to 4.6 times improvement in efficiency. The metasurface also increases the coverage area from around 5 cm by 5 cm with over 40\% efficiency to around 10 cm by 10 cm with over 70\% efficiency. We further show that the system can support multiple receivers. Besides increasing the overall efficiency, we demonstrate tuning the power division between the multiple receivers, enabling compensation of receivers of different sizes to achieve their desired power.
\end{abstract}


{}

\definecolor{limegreen}{rgb}{0.2, 0.8, 0.2}
\definecolor{forestgreen}{rgb}{0.13, 0.55, 0.13}
\definecolor{greenhtml}{rgb}{0.0, 0.5, 0.0}

\section{Introduction}

\IEEEPARstart{W}{ireless} power transfer (WPT) systems have shown their broad applications in charging mobile devices \cite{hui2013planar}, electrical vehicles \cite{li2014wireless, chen2021stability,chen2016cost}, sensing platforms \cite{sample2007design}, and implanted medical devices \cite{ho2014wireless}. Inductive WPT is the most commonly used method among radio frequency WPT \cite{ozaki2021wireless}, acoustic WPT \cite{kiziroglou2017acoustic}, and laser WPT \cite{liu2016charging}; it uses the magnetic near field to transmit power across distance \cite{zhang2016efficiency}, which has the advantages of high power volume, safety, and low radio frequency interference \cite{hui2013critical}. However, inductive WPT systems are greatly limited by the fast spatial dispersion and low penetration depth of the transmitting (Tx) coil's magnetic near field. To achieve sufficient efficiency, the receiving (Rx) coil usually requires precise alignment to the Tx coil \cite{karalis2008efficient,assawaworrarit2017robust}. Free-positioning WPT systems are proposed to eliminate such a requirement, which saves the users tremendous efforts in such alignment and also increases the success rate of charging \cite{hui2005new}. 

Free-positioning WPT systems typically employ multiple Tx coils \cite{jadidian2014magnetic}. The Tx coils are driven by independent active feeding channels to be selectively activated when they are close to the Rx coils \cite{zhang2016efficiency, wang2021comparative}. Multi-user compatibility is also an advantage of the free-positioning WPT systems \cite{wang2020robust}. Due to the increased coverage area and tunability of the power-carrying magnetic field, more than one charging device can be coupled to the WPT transmitter \cite{shi2015wireless,waters2015power}. Therefore, one can charge multiple devices simultaneously with a single wireless charging pad. However, the complexity of the system significantly increases the cost and weight of the wireless charger. Complicated power electronics also limit the number of Tx coils, which further limits the resolution of the magnetic field control and the charging efficiency. These disadvantages prevent the commercial application of such a technology, represented by the freeze of AirPower's release \cite{graham2019wireless}.

To achieve free-positioning WPT without multiple Tx coils, we employ the concept of metasurfaces. Metasurfaces are structural materials that consist of a periodic array of resonators \cite{holloway2012overview}. In our previous work, we have developed the coupled-mode theory of a novel near field metasurface capable of shaping the magnetic field on demand \cite{wang2021demand}. We have theoretically demonstrated its applications in magnetic resonance imaging \cite{wang2021demand, zhao2022ultrathin} and WPT \cite{wang2021wearable}. In this paper, we interpret the theory with a mutual induction model and extended it by developing the optimized configuration of the metasurface to maximize the efficiency. We validate the application of free-positioning and multi-user supportive WPT experimentally using the Qi standard, which is the most commonly used WPT protocol for mobile devices \cite{hui2013planar}. Compared to a single Tx coil WPT system, the metasurface significantly increases the peak efficiency from 48.2\% to 77.5\% and enables a near-flat efficiency distribution across around 10 cm by 10 cm area with an efficiency of over 70\%. The metasurface WPT system supports multiple receivers and precisely controls their power division by changing the reactance distribution. We further show the enhanced efficiency in charging multiple receivers with different sizes, a significant challenge in traditional designs to charge the smaller receiver. Such a size difference can be compensated by the metasurface to achieve similar power delivery.

\section{Theoretical Analysis}
\subsection{Mutual Coupling Model of Metasurface-Enhanced WPT system}
The metasurface-enhanced WPT system uses the metasurface to reform the magnetic field. As shown in Fig. 1, the metasurface consists of an array of unit cells being strongly coupled to each other. The metasurface is coupled to the Tx coil and the Rx coil, serving as a highly tunable passive relay between the transmitting side and the receiving side. To describe such a system and calculate the optimal reactance configuration of the metasurface for maximum transmission efficiency, we developed the mutual induction model of the system.
\begin{align}
\left[ {\begin{array}{*{20}{c}}
  {{V_{Tx}}} \\ 
  {{{\mathbf{V}}_{\mathbf{u}}}} \\ 
  {{{\mathbf{V}}_{{\mathbf{Rx}}}}} 
\end{array}} \right] = j\omega \left[ {\begin{array}{*{20}{c}}
  {{L_{Tx}}}&{{{\mathbf{M}}_{{\mathbf{Tu}}}}}&{{{\mathbf{M}}_{{\mathbf{TR}}}}} \\ 
  {{{\mathbf{M}}_{{\mathbf{uT}}}}}&{{{\mathbf{L}}_{{\mathbf{uu}}}}}&{{{\mathbf{M}}_{{\mathbf{uR}}}}} \\ 
  {{{\mathbf{M}}_{{\mathbf{RT}}}}}&{{{\mathbf{M}}_{{\mathbf{Ru}}}}}&{{{\mathbf{L}}_{{\mathbf{Rx}}}}} 
\end{array}} \right] \cdot \left[ {\begin{array}{*{20}{c}}
  {{I_{Tx}}} \\ 
  {{{\mathbf{I}}_{\mathbf{u}}}} \\ 
  {{{\mathbf{I}}_{{\mathbf{Rx}}}}} 
\end{array}} \right]
\end{align}
where italic characters represent scalers, and bold characters represent vectors or matrices.$[{V_{Tx}}
$, ${{\mathbf{V}}_{\mathbf{u}}}$, and ${{\mathbf{V}}_{{\mathbf{Rx}}}}$ are the voltage of the Tx coil, the unit cells, and the Rx coil(s); $\omega  = 2\pi f$, $f$ is the operational frequency; ${L_{Tx}}$, ${{\mathbf{L}}_{{\mathbf{uu}}}}$, and ${{\mathbf{L}}_{{\mathbf{Rx}}}}$ are the self-inductance of the Tx coil, the unit cells, and the Rx coil(s), respectively; ${{\mathbf{M}}_{{\mathbf{Tu}}}}$, ${{\mathbf{M}}_{{\mathbf{TR}}}}$, and ${{\mathbf{M}}_{{\mathbf{Ru}}}}$ are the mutual inductance between the Tx coil and the unit cells, the Tx coil and the Rx coil, and the Rx and the unit cells. Due to the principle of reciprocity, ${{\mathbf{M}}_{{\mathbf{uT}}}} = {{\mathbf{M}}_{{\mathbf{Tu}}}}$, ${{\mathbf{M}}_{{\mathbf{RT}}}} = {{\mathbf{M}}_{{\mathbf{TR}}}}$, and ${{\mathbf{M}}_{{\mathbf{uR}}}} = {{\mathbf{M}}_{{\mathbf{Ru}}}}$. If the metasurface has ${N_u}$ unit cells, and there are ${N_{Rx}}$ Rx coils, ${{\mathbf{V}}_{\mathbf{u}}}$ will be a ${N_u} \times 1$ vector; ${{\mathbf{V}}_{{\mathbf{Rx}}}}$ will be ${N_{Rx}} \times 1$ vector; ${{\mathbf{L}}_{{\mathbf{uu}}}}$ will be ${N_u} \times {N_u}$ matrix; ${{\mathbf{L}}_{{\mathbf{Rx}}}}$ will be ${N_{Rx}} \times {N_{Rx}}$ matrix; ${{\mathbf{M}}_{{\mathbf{Tu}}}}$ will be $1 \times {N_u}$ vector; ${{\mathbf{M}}_{{\mathbf{TR}}}}$ will be $1 \times {N_{Rx}}$ vector; ${{\mathbf{M}}_{{\mathbf{Ru}}}}$ will be ${N_{Rx}} \times {N_u}$ matrix.
\begin{figure}[!t]\centering
	\includegraphics[width=7.5cm]{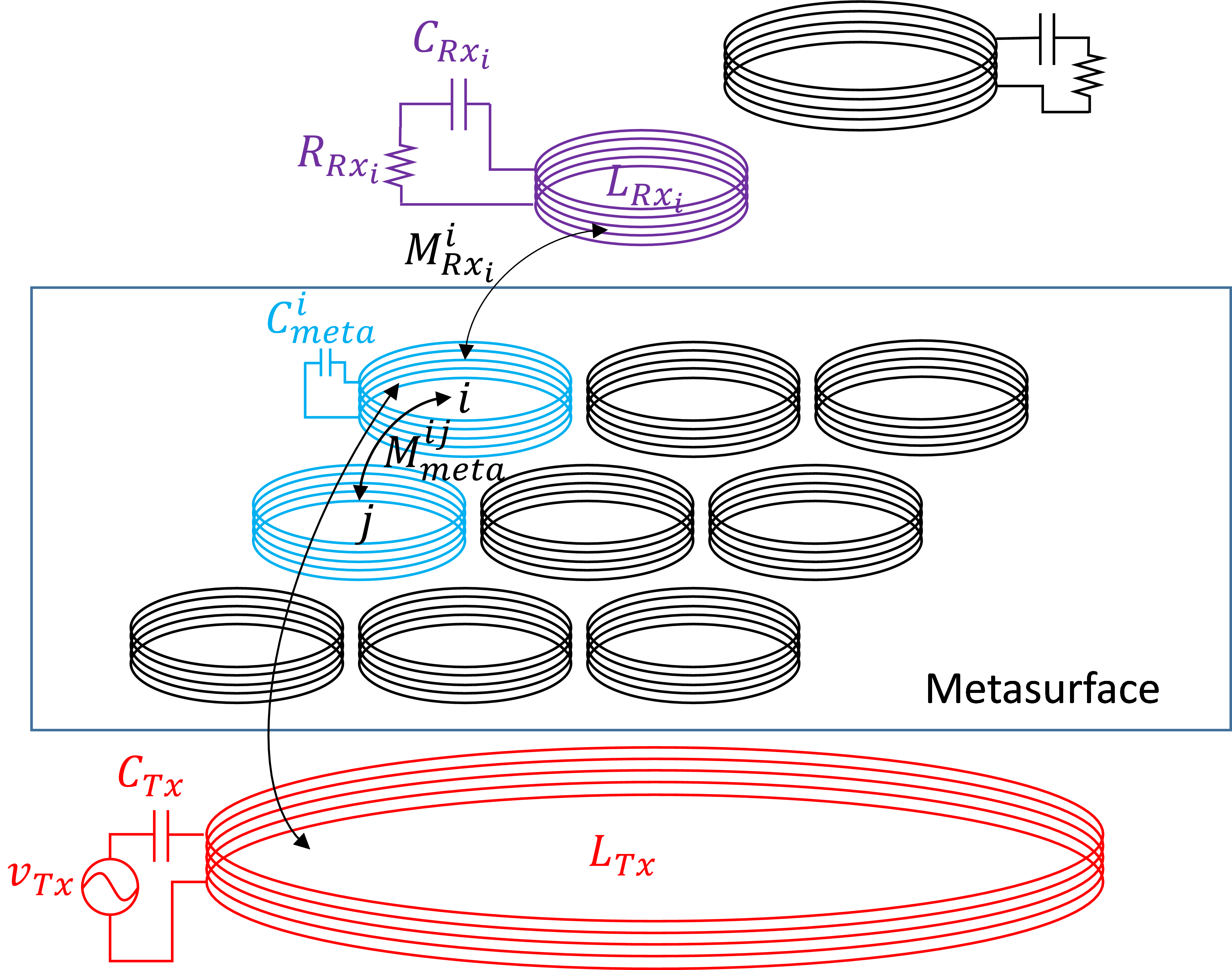}
	\caption{Circuit diagram of a metasurface-enhanced wireless power transfer (WPT) system with multiple receivers.}\label{FIG_1}
\end{figure}

The unit cells are closed-loop with only a compensation capacitor in series, thus, the voltage of the unit cells is
\begin{align}
{{\mathbf{V}}_{\mathbf{u}}} = {{\mathbf{Z}}_{\mathbf{u}}}{{\mathbf{I}}_{\mathbf{u}}}
\end{align}
where ${{\mathbf{Z}}_{\mathbf{u}}} = {R_u}{\mathbf{I}} + j\left[ {\begin{array}{*{20}{c}}
  {{X_1}}&0&0 \\ 
  0& \ddots &0 \\ 
  0&0&{{X_{{N_u}}}} 
\end{array}} \right]$
, ${R_u}$ is the internal resistance of the unit cell, which is measured to be around 0.075 $\Omega$; ${\mathbf{I}}$ is an ${N_u} \times {N_u}$ identity matrix; ${X_i}$ is the reactance of the i-th unit cell, ${X_i} = \omega {L_u} - \frac{1}{{\omega {C_{ui}}}}$; ${L_u}$ is the self-inductance of the unit cell, which is measured to be 1.77 $\mu H$; and ${C_{ui}}$ is the compensation capacitance of each unit cell. The coils of the unit cells are identical, therefore, have the same self-inductance and internal resistance. It is important to note that the compensation capacitance of each unit cell, ${C_{ui}}$, is chosen specifically to control reactance distribution of the metasurface, and ultimately, the current distribution of the unit cells and the magnetic field distribution reformed by the metasurface. 

\subsection{Physical Configuration of the Metasurface}
Fig. 2 shows the schematic of the metasurface, which consists of unit cells in a face-centered-cubic structure with two layers to ensure a strong coupling between the adjacent unit cells. The z-separation between the two layers is 0.75 cm. For the unit cell, the diameter is 4 cm; the thickness is 0.75 cm; the periodicity is 3 cm, which is the same in both the x and y directions. Note that the configuration of the metasurface is not unique but can be chosen according to the application. The size, periodicity, and number of unit cells can be adjusted according to the required coverage area and Rx coil size. Specifically, more unit cells can be applied to increase the coverage area; the unit cells should have a similar size to the Rx coil to provide a high enough field shaping resolution while keeping a sufficiently low power consumption of the unit cells.
\begin{figure}[!t]\centering
	\includegraphics[width=6cm]{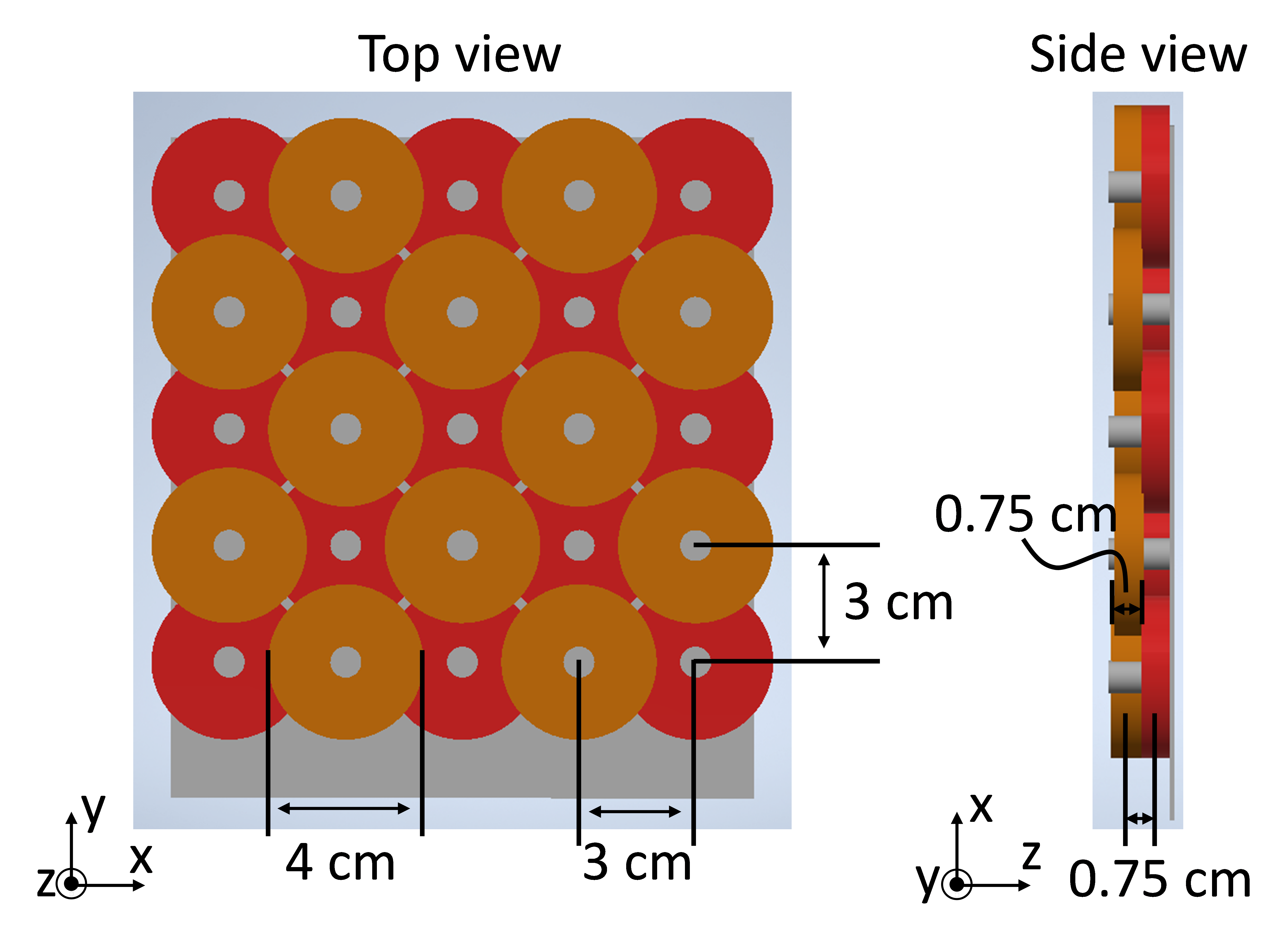}
	\caption{Schematics of the physical configuration of the metasurface.}\label{FIG_2}
\end{figure}

\subsection{Inverse Design of the Metasurface's Reactance Configuration}
In the following, we introduce how to back-calculate the required reactance distribution given a targeted current distribution of the unit cells and the physical configuration of the metasurface. When there is no receiver, we combine Eq. (1) and Eq. (2) to get the linear equation for solving the current distribution of the unit cells as ${{\mathbf{Z}}_{\mathbf{u}}}{{\mathbf{I}}_{\mathbf{u}}} = j\omega {\mathbf{M}}{}_{{\mathbf{uT}}}{{\mathbf{I}}_{{\mathbf{Tx}}}} + j\omega {{\mathbf{L}}_{{\mathbf{uu}}}}{{\mathbf{I}}_{\mathbf{u}}}$. Therefore, we can solve the current distribution of the unit cells, i.e., the metasurface's mode, as 
\begin{align}
{{\mathbf{I}}_{\mathbf{u}}} = j\omega {({{\mathbf{Z}}_{\mathbf{u}}} - j\omega {{\mathbf{L}}_{{\mathbf{uu}}}})^{ - 1}}{\mathbf{M}}{}_{{\mathbf{uT}}}{{\mathbf{I}}_{{\mathbf{Tx}}}}
\end{align}
We set the targeted current distribution of the unit cells as ${{\mathbf{I}}_{{\mathbf{u - tar}}}}$. By setting the resulting current distribution ${{\mathbf{I}}_{\mathbf{u}}}$ in Eq. (3) as the targeted current distribution ${{\mathbf{I}}_{{\mathbf{u - tar}}}}$, we get the solution for the impedance distribution as
\begin{align}
{{\mathbf{Z}}_{\mathbf{u}}}{{\mathbf{I}}_{{\mathbf{u - tar}}}} = j\omega ({{\mathbf{M}}_{{\mathbf{uT}}}}{I_{Tx}} + {{\mathbf{L}}_{{\mathbf{uu}}}}{{\mathbf{I}}_{{\mathbf{u - tar}}}})
\end{align}
When the unit cells are strongly coupled, the current distribution is formed mainly by the mutual induction between the unit cells instead of the external feeding field. In such a case, the first term to the right of the equation, ${{\mathbf{M}}_{{\mathbf{uT}}}}{I_{Tx}}$, given by the mutual induction between the Tx coil and the unit cells, is significantly weaker than the second term, ${{\mathbf{L}}_{{\mathbf{uu}}}}{{\mathbf{I}}_{{\mathbf{u - tar}}}}$, given by the mutual inductance between the unit cells (especially the neighboring unit cells). Under the strong-coupling condition, the solution of the impedance distribution becomes very simplified and independent of the mutual inductance of the Tx coils to the unit cells.
\begin{align}
Z_u^j = j\omega \frac{{\sum\limits_{i = 1}^{{N_u},i \ne j} {L_{uu}^{ij}I_{u - tar}^i} }}{{I_{u - tar}^j}}
\end{align}
where $Z_u^j$ represents the impedance of the j-th unit cell, $L_{uu}^{ij}$ represents the mutual inductance between the i-th and the j-th unit cell, and $I_{u - tar}^i$ represents the targeted current of the i-th unit cell. When the internal resistance of the unit cell is small, i.e., ${R_u} \to 0$, such a relationship can be constructed by configuring the capacitance in a way that the reactance of the unit cells, ${X_u}$, equals the imaginary part of the targeted impedance distribution calculated by Eq. (5), i.e.,
\begin{align}
\omega {L_u} - \frac{1}{{\omega {C_{uj}}}} = \omega \frac{{\sum\limits_{i = 1}^{{N_u},i \ne j} {L_{uu}^iI_{u - tar}^i} }}{{I_{u - tar}^j}}
\end{align}

We use an analytical approach instead of numerical simulations to speed up the calculation while solving the mutual inductance between the Tx coil, the unit cells, and the Rx coil. The mutual inductance is given by Neumann's formula 
\begin{align}
{M_{12}} = \frac{{{\mu _0}}}{{4\pi }}\oint\limits_{\partial {\Omega _1}} {\oint\limits_{\partial {\Omega _2}} {\frac{1}{{{r_{12}}}}d{l_1}d{l_2}} } 
\end{align}
where $\partial {\Omega _1}$ and $\partial {\Omega _2}$ are the contours of the two coils, $d{l_1}$ and $d{l_2}$ are infinitesimal segments on the two coils, and ${r_{12}}$ is the distance between the two infinitesimal segments.

In summary, to form the targeted current distribution, we need to choose the compensation capacitors of each unit cell to form the reactance distribution as calculated by Eq. (6). To achieve an accurate resulting current distribution, the metasurface needs to follow two criteria: First, the unit cells need to be strongly coupled. The electrical potential raised by the mutual induction between the unit cells, especially the neighboring unit cells, needs to be significantly larger than the electrical potential raised by the Tx coil. Second, the unit cells need to have relatively low internal resistance.

\subsection{Optimization of the Targeted Current distribution}
The end goal of the metasurface-enhanced wireless power transfer system is to achieve high efficiency. As we have shown in our previous paper \cite{wang2021wearable}, the proper choice of the targeted current distribution can result in a highly controllable reformed magnetic field, and therefore, an enhanced efficiency. In this section, we extend this method by deriving the theoretically optimized targeted current distribution with a given Rx configuration.

In a single Rx case, when ignoring the radiation loss, the efficiency of the metasurface-enhanced WPT system is
\begin{align}
\eta  = \frac{{{{\left| {{I_{Rx}}} \right|}^2}{R_L}}}{{{{\left| {{I_{Rx}}} \right|}^2}({R_L} + {R_{Rx}}) + {{\left| {{{\mathbf{I}}_{\mathbf{u}}}} \right|}^2}{R_u} + {{\left| {{I_{Tx}}} \right|}^2}{R_{Tx}}}}
\end{align}
where ${R_{Rx}}$ and ${R_{Tx}}$ are the internal resistance of the Rx and Tx coils. When the metasurface is configured according to the targeted current distribution and follows the two criteria as discussed in the previous section, ${\mathbf{|}}{{\mathbf{I}}_{\mathbf{u}}}{\mathbf{| = |}}{{\mathbf{I}}_{{\mathbf{u - tar}}}}{\mathbf{|}}$.

According to Eq. (1), under perfect compensation of the Rx coil, the Rx coil current can be calculated as ${I_{Rx}} = \frac{{j\omega ({M_{RT}}{I_{Tx}} + {{\mathbf{M}}_{{\mathbf{Ru}}}}{{\mathbf{I}}_{{\mathbf{u - tar}}}})}}{{{R_L} + {R_{Rx}}}}$. Substituting it into Eq. (8), the efficiency becomes
\begin{align}
\eta  = \frac{K}{{K + {{\left| {{{\mathbf{I}}_{{\mathbf{u - tar}}}}} \right|}^2}{R_u} + {{\left| {{I_{Tx}}} \right|}^2}{R_{Tx}}}}
\end{align}
where $K = {\omega ^2}{({M_{RT}}{I_{Tx}} + {{\mathbf{M}}_{{\mathbf{Ru}}}}{{\mathbf{I}}_{{\mathbf{u - tar}}}})^2}\frac{{{R_L}}}{{{{({R_L} + {R_{Rx}})}^2}}}$.
The Jacobian matrix of the efficiency over the targeted current distribution yields the condition for the maximum efficiency. 
\begin{align}
\frac{{\partial \eta }}{{\partial {{\mathbf{I}}_{\mathbf{u}}}}} = 0
\end{align}
Solving Eq. (10), the optimal targeted current distribution is given by
\begin{align}
{{\mathbf{I}}_{{\mathbf{u - tar}}}} = c{{\mathbf{M}}_{{\mathbf{uR}}}}
\end{align}
where $c$  is a constant related to ${R_L}$, ${R_{Rx}}$, ${R_{Tx}}$, ${{\mathbf{M}}_{{\mathbf{Ru}}}}$, and ${M_{RT}}$. The capacitance distribution given by Eq. (6) is independent of the absolution amplitude but normalized distribution of the targeted current distribution. Therefore, the constant   does not matter. This result fits with the active control algorithm of the coil array for WPT \cite{zhang2016efficiency}. Unlike these studies, we do not require any active components to form such a current distribution.

\section{PRACTICAL IMPLEMENTATION AND EVALUATION}
\subsection{Experiment Setup}
The overall experiment setup, including the driving and characterization system, is shown in Fig. 3(a) and the metasurface-enhanced WPT system are shown in the dashed box of Fig. 3(a) and Fig. 3(b). The external characterization system contains a waveform generator (Keysight 33500B Waveform Generator), an oscilloscope (Agilent DSO-x 3034A), and a controlling computer. As shown in Fig. 3(a), we use the waveform generator as an AC power supply to the Tx coil. The operational frequency of the waveform generator is under the frequency range of the Qi standard, i.e., 100 kHz$\sim$200 kHz. We use the oscilloscope to monitor the voltage and the current of the Tx and Rx coils. Both the waveform generator and the oscilloscope are controlled by MATLAB R2022a through the National Instrument VISA 2022Q3 interface on the computer. 

We use MATLAB to control the frequency of the feeding voltage and read the voltage in the time domain of the Tx coil and the Rx coil (both in series with compensation capacitors) and the voltage of their compensation capacitors (Fig. 3(b)). The complex voltages are calculated by taking Fast Fourier transform (FFT) of the voltage signal in the time domain multiplied with a Hanning window. We calculate the complex voltages of the Tx coil, Rx coil, and their compensation capacitors. The complex currents of the Tx and Rx coils are measured through the complex voltages of their compensation capacitors as $I = V \cdot (j\omega C)$.

Fig. 3(b) shows the metasurface-enhanced WPT system, including a Tx coil, a metasurface, and an Rx coil. Please note that the separation in the figure does not represent the practical working condition but is increased to better show the three components. In the practical setup, the center of the metasurface is placed 3 cm from the Tx coil and 3 cm from the Rx coil along the z-axis. The center of the metasurface is aligned with the center of the Tx coil in the x-y plane. The x-y position of the receiver is adjustable, and multiple receivers can be implemented. The metasurface can be removed from the system while keeping the position of the Tx and Rx coils to measure the magnetic power density and efficiency without the metasurface. 

The magnetic probe, Rx coil, and unit cells of the metasurface are shown in Fig. 3(c). The magnetic probe is an open coil. Both the Tx and Rx coils are serial-compensated at the operational frequency of 190 kHz and have 10 turns. The Tx coil has a diameter of 10 cm and the Rx coil has a diameter of 4 cm. Both the Tx coil and the Rx coil are serial-compensated at the operational frequency. The Rx coil is in series to a load with a resistance of 1 $\Omega$. The unit cells are tuned to different reactance at the operational frequency. Specifically, we use the compensation capacitances of 410 nF, 480 nF, 580 nF, and 1130 nF to achieve the reactance of 0 $\Omega$, 0.3 $\Omega$, 0.6 $\Omega$, and 1.3 $\Omega$, respectively. And we short the coil to achieve the reactance of 1.8 $\Omega$. All the coils are wired with No. 20 American wire gauge (AWG) Litz wires. The inductance of all the coils and the capacitance of all the compensation capacitors are measured with an LCR meter (HP 4284A Precision LCR Meter). 
\begin{figure}[!t]\centering
	\includegraphics[width=8.5cm]{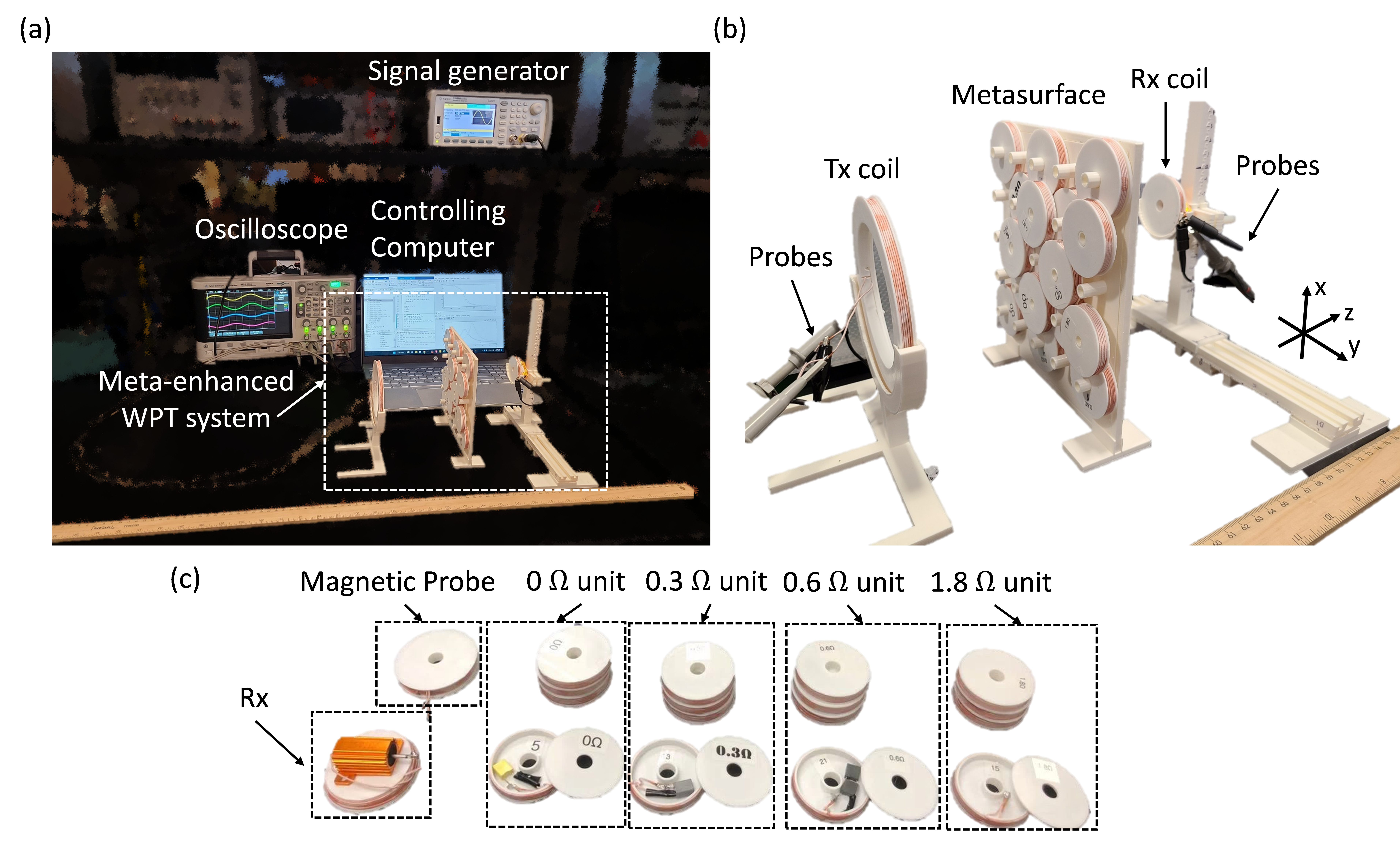}
	\caption{Photographs of the experiment setup. (a) Overall setup. (b) Metasurface-enhanced WPT system. (c) Magnetic probe, receiving (Rx) coil, and unit cells with different reactance.}\label{FIG_4}
\end{figure}


\subsection{Mutual Coupling Between the Unit Cells}
The self-inductance matrix ${{\mathbf{L}}_{{\mathbf{uu}}}}$ consists of mutual inductance between all the unit cells, ${{\mathbf{L}}_{{\mathbf{uu}}}} = \left[ {\begin{array}{*{20}{c}}
  {{L_1}}& \ldots &{{M_{1{N_u}}}} \\ 
   \vdots & \ddots & \vdots  \\ 
  {{M_{{N_u}1}}}& \cdots &{{L_{{N_u}}}} 
\end{array}} \right]$, and is critical in solving the reactance distribution in Eq. (6). The mutual inductance is decided by the physical displacement of the unit cells. The metasurface consists of unit cells of two layers, and we calculate them separately through Eq. (7). To measure the mutual inductance, we first measure the two-port impedance between two unit cells, i.e., the open circuit voltage on the second unit cell divided by the current of the first unit cell, ${Z_{12}} = \frac{{{V_2}}}{{I{}_1}}$. The mutual inductance between the two coils is then calculated by ${M_{12}} = \frac{{{Z_{12}}}}{{j\omega }}$. As shown in Fig. 4, the theoretical mutual inductance for both the same layer (Fig. 4(a)) and the different layers (Fig. 4(b)) fit with the measurements, which guarantees the accuracy of the self-induction matrix ${{\mathbf{L}}_{{\mathbf{uu}}}}$.
\begin{figure}[!t]\centering
	\includegraphics[width=9.5cm]{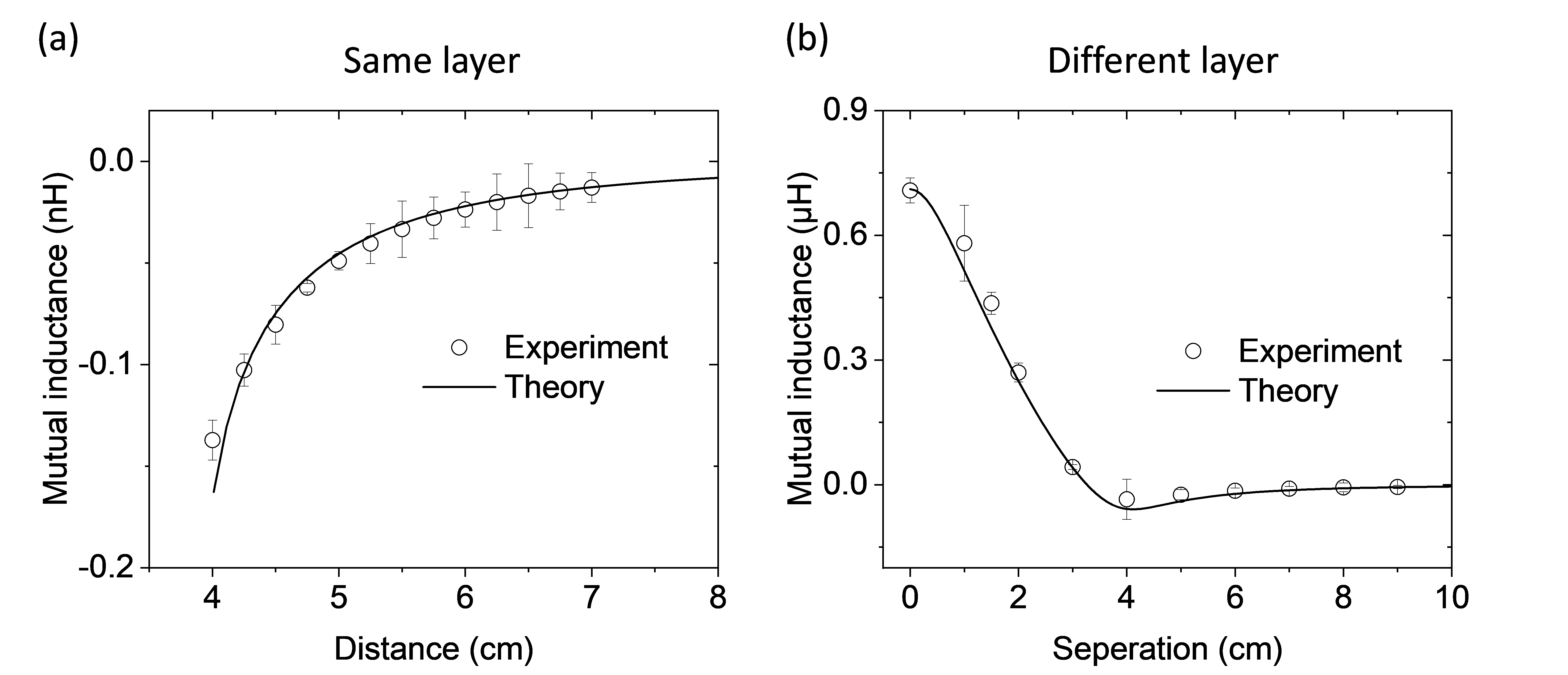}
	\caption{Mutual inductance between the unit cells as a function of their separation in (a) the same layer and (b) the different layers.}\label{FIG_7}
\end{figure}
\subsection{Efficiency Measurement}
To measure the efficiency, we take the ratio between the output power of the loading resistor to the input power of the Tx coil
\begin{align}
\eta  = \frac{{{V_L}^2/{R_L}}}{{\operatorname{Re} ({V_{Tx}}^* \cdot {I_{Tx}}))}}
\end{align}
where ${V_L}$ is the voltage of the loading resistor, ${R_L}$ is the load resistance, being 1 $\Omega$ in the experiment, ${V_{Tx}}$ is the complex voltage of the Tx coil, ${I_{Tx}}$ is the complex current of the Tx coil. The current can be measured by the voltage of the compensation capacitor of the Tx coil as ${I_{Tx}} = j\omega {C_{Tx}}{V_C}$, where ${C_{Tx}}$ is the compensation capacitance, being 28 nF in the experiment, and ${V_C}$ is the measured complex voltage of the capacitor.

\section{COMPARISON OF THEORETICAL AND PRACTICAL RESULTS}
The general optimization of the metasurface-enhanced WPT system follows three steps: (1) Calculate the targeted current distribution given by the mutual inductance distribution ${{\mathbf{M}}_{{\mathbf{uR}}}}$ (Eq. (11)); (2) Calculate the reactance distribution with Eq. (6) using the targeted current distribution from step (1); (3) Choose appropriate unit cells to form the reactance distribution. 
\subsection{Magnetic Field Shaping}
When the Rx coil is placed with an offset of one period (3 cm) to the center of the metasurface in both the x and y directions, the solved optimized current distribution with Eq. (11) is shown in Fig. 5(a). We use Eq. (6) to calculate the targeted reactance distribution based on the optimized targeted current distribution. In practice, due to the limited option of the unit cell's reactance, we discretize the continuous reactance distribution into the digits of 0 $\Omega$, 0.3 $\Omega$, 0.6 $\Omega$, 1.3 $\Omega$, and 1.8 $\Omega$. For the region with the targeted normalized current distribution lower than 0.1, we switched off the unit cells by removing them from the region. The reactance distribution is shown in Fig. 5(b). Note that in WPT, the accuracy of the magnetic field shaping does not significantly influence the efficiency. Therefore, we do not need a lot of digits of the unit cell to achieve the claimed functions. In some scenarios, even single-digit switching can achieve a substantial effect \cite{ranaweera2019active, shi2016large}. However, the more complicated tasks that we introduce in later text such as tuning the power division require a few more digits of the unit cell's reactance. 
\begin{figure}[!t]\centering
	\includegraphics[width=7cm]{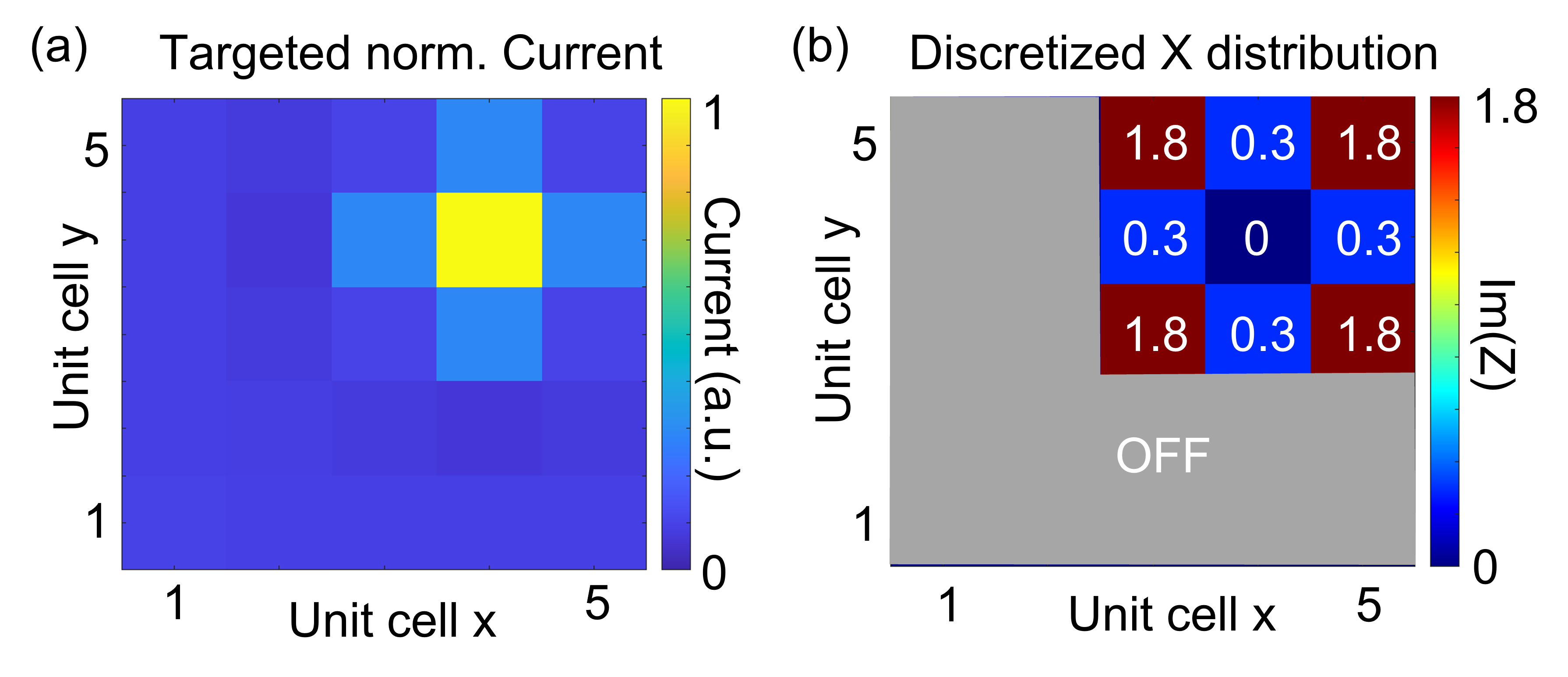}
	\caption{Theoretical optimized reactance distribution. (a) Targeted current distribution by Eq. (11). (b) Reactance distribution by Eq. (6).}\label{FIG_9}
\end{figure}
We measure the magnetic power density with an open Rx coil. The magnetic power density is,
\begin{align}
{U_B} = \frac{1}{{2{\mu _0}}}{\left| B \right|^2}
\end{align}
where ${\mu _0}$ is vacuum permeability. The magnetic flux density can be measured by the voltage of the open Rx coil, ${V_{Rx - open}}$, as
\begin{align}
B = \frac{{{V_{Rx - open}}}}{{j\omega n\pi {r^2}}}
\end{align}
where $n$ and $r$ are the number of turns and the radius of the Rx coil.

The magnetic power density measures the magnetic field reformed by the metasurface. As shown in Fig. 6, the metasurface is able to reshape the magnetic field on-demand, resulting in a more confined magnetic power density distribution with an increased peak intensity (Fig. 6(b)) compared to the case without the metasurface (Fig. 6(a)). The measured magnetic power density distribution confirms this conclusion (Fig. 6(c) and 6(d)). Note that the magnetic power density does not directly correspond to the power that the Rx coil receives with a loading resistor. As when the Rx coil is not an open circuit, its current will generate potential to the metasurface and the Tx coil, and result in a reduced current of both the metasurface and the Tx coil. Therefore, the received magnetic power density will be reduced. The magnetic power density here is to confirm the magnetic field-shaping capability of the metasurface without an external load.

\begin{figure}[!t]\centering
	\includegraphics[width=7cm]{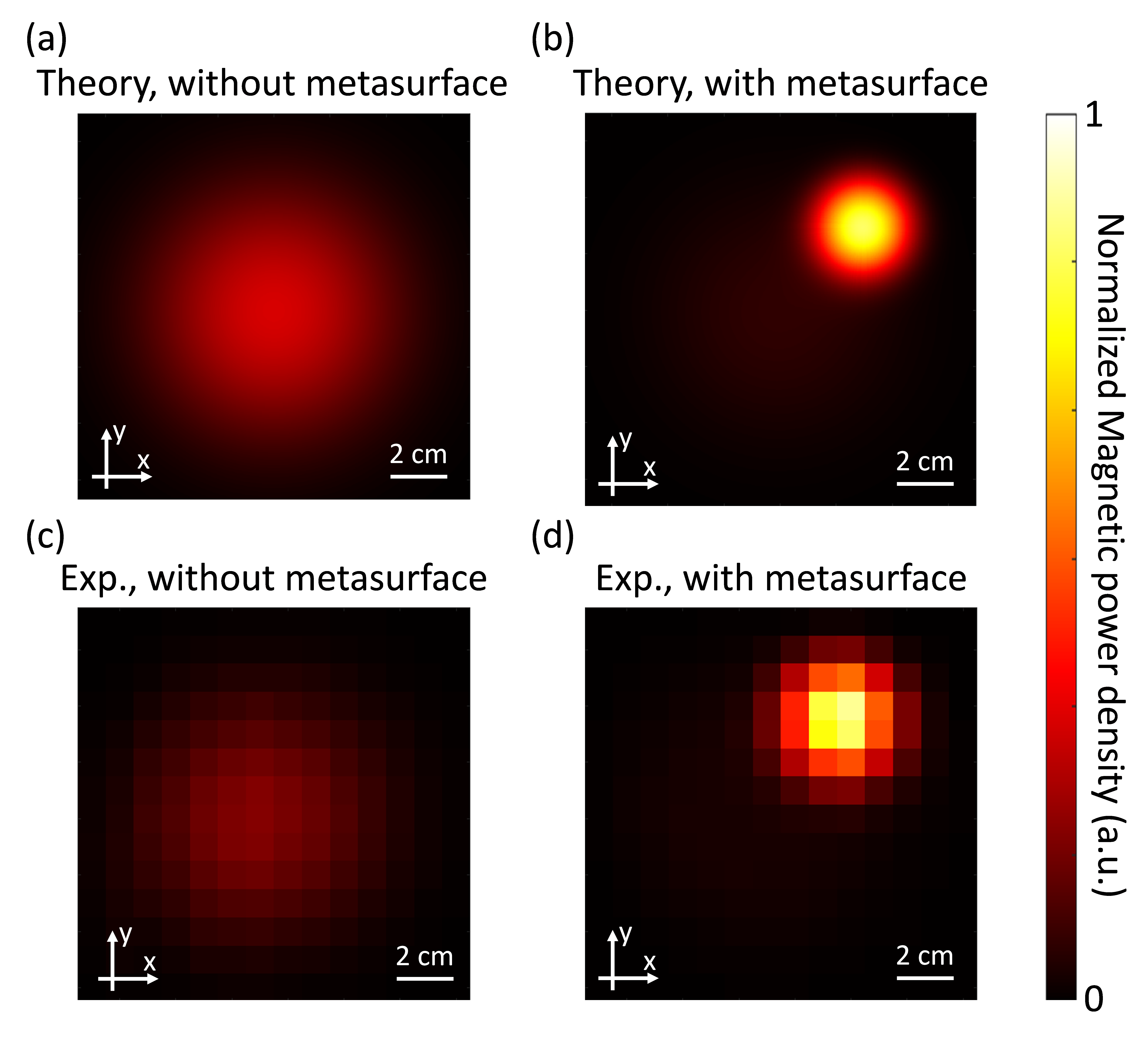}
	\caption{Spectrum of the magnetic power density with the Rx coil at x = 3 cm, and y = 3 cm, i.e., the peak location of the targeted current distribution.}\label{FIG_10}
\end{figure}

\subsection{Efficiency with a Single Rx Coil}
The metasurface can redistribute the magnetic flux and enhance the efficiency at selected positions. We choose the targeted location to be offset from the center of the metasurface by 1 period (3 cm) in both the x and y directions, the same as we discussed in Fig. 6. We calculate the theoretical efficiency with Eq. (9). The efficiency distributions without and with the metasurface are shown in Fig. 7(a) and Fig. 7(b). The peak efficiency is 38.1\% without the metasurface and increases to 67.27\% with the metasurface. Furthermore, the peak position shifts from the center to the targeted region with 3 cm offset in both the x and y directions.

We measure the efficiency map by scanning the Rx coil's x-y position and calculate the efficiency by Eq. (14). Fig. 7(c) shows the measured efficiency without the metasurface, which fits with the theoretical result in Fig. 7(a). The peak efficiency in this configuration is 35.4\% and occurs at the center position. Fig. 7(d) shows the measured efficiency with the metasurface, which produces a peak efficiency of 63.5\% at the position of x = 3 cm and y = 3 cm. 

\begin{figure}[!t]\centering
	\includegraphics[width=7cm]{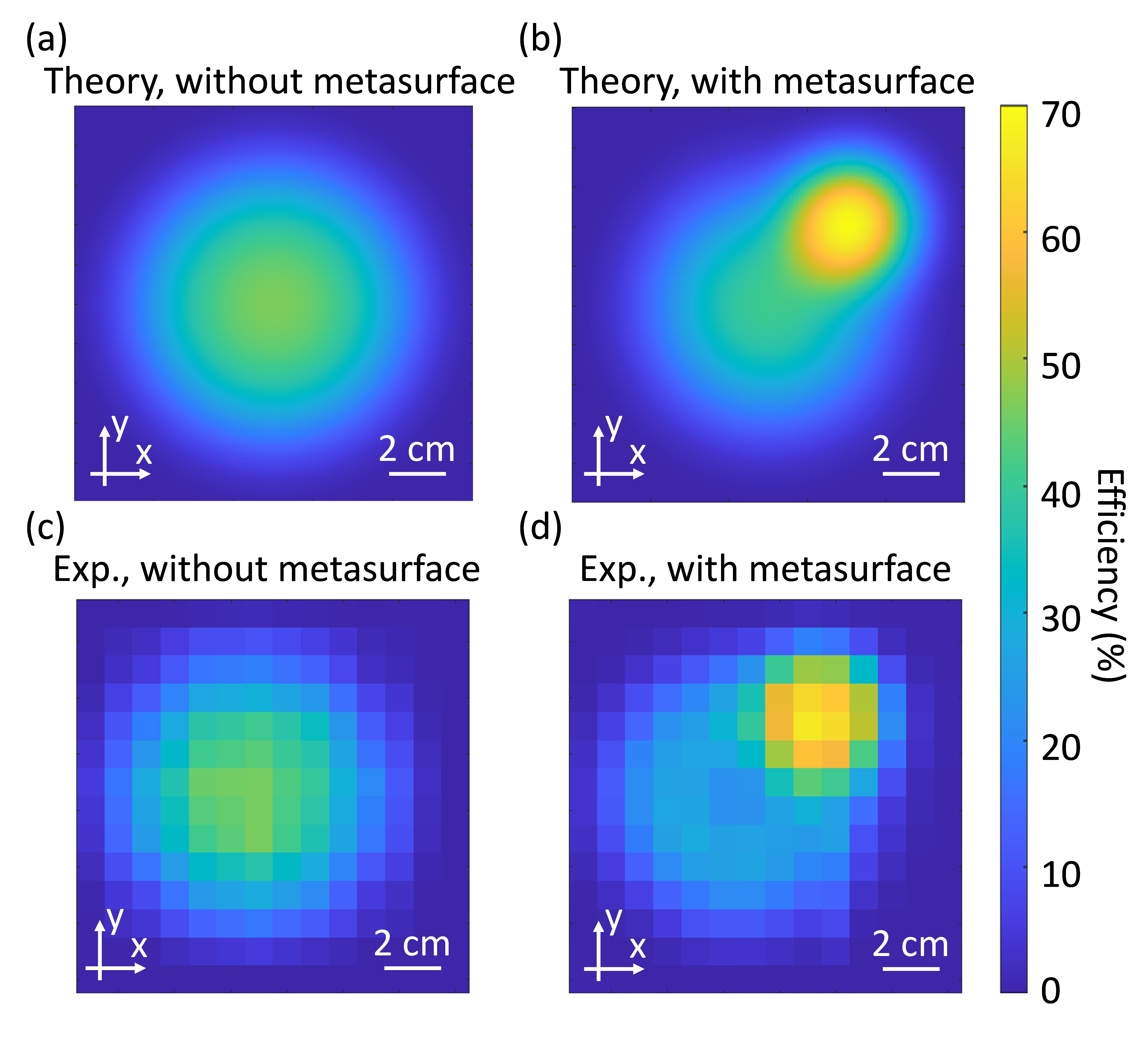}
	\caption{Transmission efficiency of a single receiver enhanced by the metasurface. Theoretical efficiency (a) without and (b) with the metasurface. Measured efficiency (c) without and (d) with the metasurface.}\label{FIG_12}
\end{figure}

\subsection{Free-Positioning WPT}
We can tune the position of the metasurface-formed magnetic flux beam by reconfiguring the center positioning of the targeted current distribution, and therefore, always target the magnetic flux beam to the Rx coil's position to enable free-positioning WPT. Fig. 8(a) shows the center positions of three targeted current distributions. The centers are tuned to be offset by -3 cm, 0 cm, and 3 cm in the x direction, respectively, and centered in the y direction. These three targeted current distributions are named as mode 1, mode 2, and mode 3. We measured the efficiency distribution along the x direction with and without the metasurface for these three targeted current distributions. As shown in Fig. 8(b), when tuning the center of the magnetic flux beam, the peak position of the efficiency moves accordingly.

The peak efficiency of all the three modes can reach over 75\%. Particularly, compared to the efficiency without the metasurface, where the maximum efficiency can only achieve 48.2\% at the center, the metasurface significantly boosts the efficiency to 77.5\% with mode 2. Moreover, due to the metasurface's ability in redirecting magnetic flux to an off-centered position, we improve the efficiency from 35.8\% to 74.5\% with mode 1 and from 35.0\% to 71.1\% with mode 3.
\begin{figure}[!t]\centering
	\includegraphics[width=8.5cm]{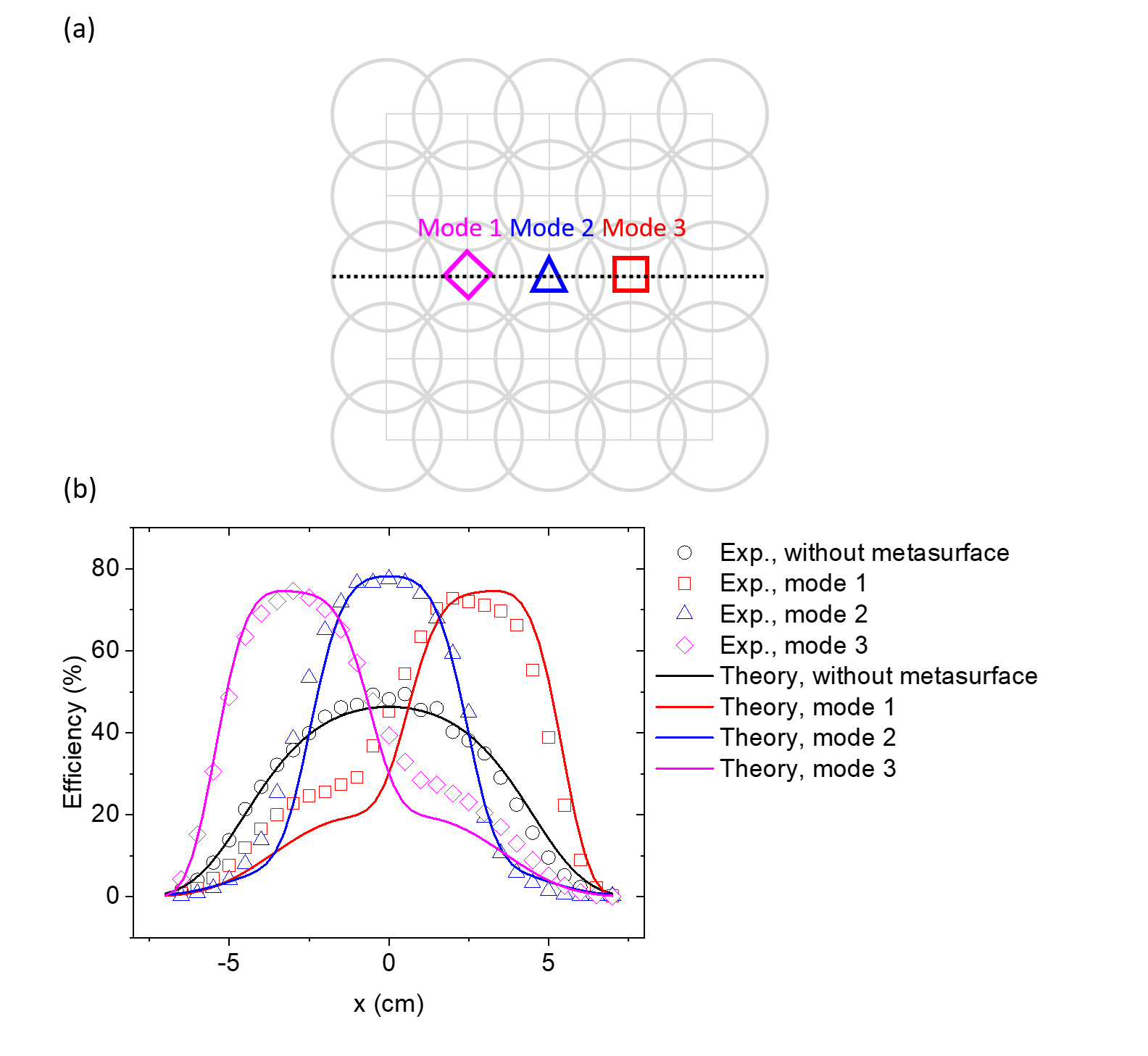}
	\caption{Efficiency distribution of different targeted current distribution. (a) Peak positions of the three targeted current distribution. (b) Theoretical and measured efficiency distribution in the x direction along the dashed line in (a).}\label{FIG_14}
\end{figure}
As shown in Fig. 9, by switching the targeted current distribution and choosing the one with optimized efficiency, we get an efficiency of around 75\% across a distance of around 10 cm. Compared to the efficiency distribution without the metasurface, the metasurface can provide an efficiency improvement of up to 4.6 times. Note that although we change the configuration manually in this paper, the principle can be easily extended to reconfigurable metasurfaces with digitally switchable unit cells. A near-flat efficiency distribution can be obtained at any location within the metasurface's coverage region. Therefore, a free-positioning WPT system is enabled by the metasurface.
\begin{figure}[!t]\centering
	\includegraphics[width=7cm]{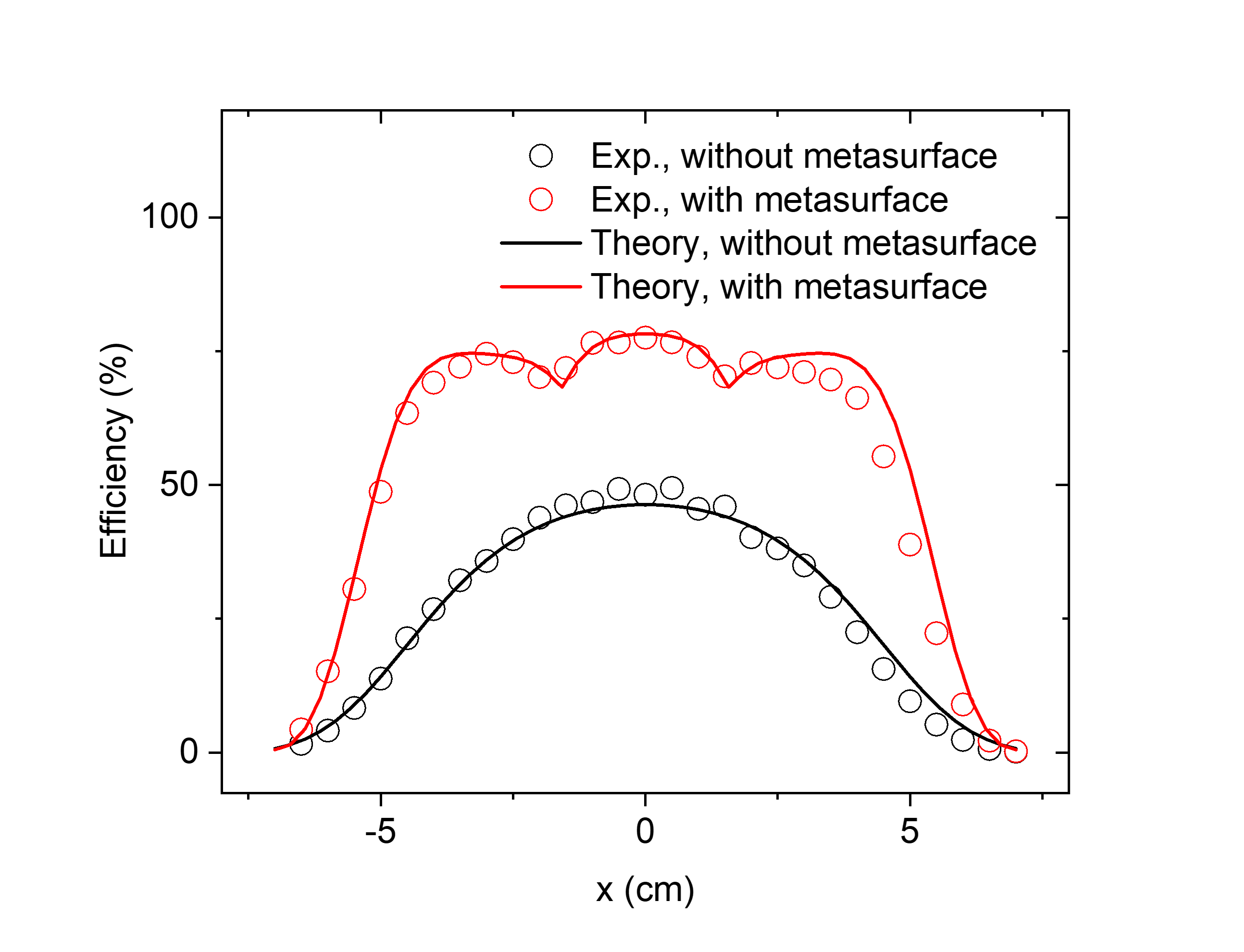}
	\caption{Free-positioning enhancement enabled by the metasurface.}\label{FIG_15}
\end{figure}

\subsection{Multiple Magnetic Flux Beams with Tunable Intensity Ratio}
We can configure the metasurface to form more than one magnetic flux beam. The targeted current distribution is summed over two magnetic flux beams centered at x = 3 cm, y = 3 cm, and x = -3 cm, y = -3 cm with intensity ratios of 1:1 and 2:1. Similar to the single receiver case, the reactance distribution can be calculated by Eq. (6) and discretized according to the reactance options of the unit cells. The results are shown in Fig. 10(a) and 10(b). The measured magnetic power density distributions in Fig. 10(c) and 10(d) confirm the that the metasurface can redirect the magnetic flux into two beams with similar intensities and non-equal intensities. This unique advantage is given by the accurate field-shaping capability of our proposed metasurface and can be used to control the power division between multiple Rx coils and compensate the power difference given by non-uniformed coupling to the Tx coil.

\begin{figure}[!t]\centering
	\includegraphics[width=8cm]{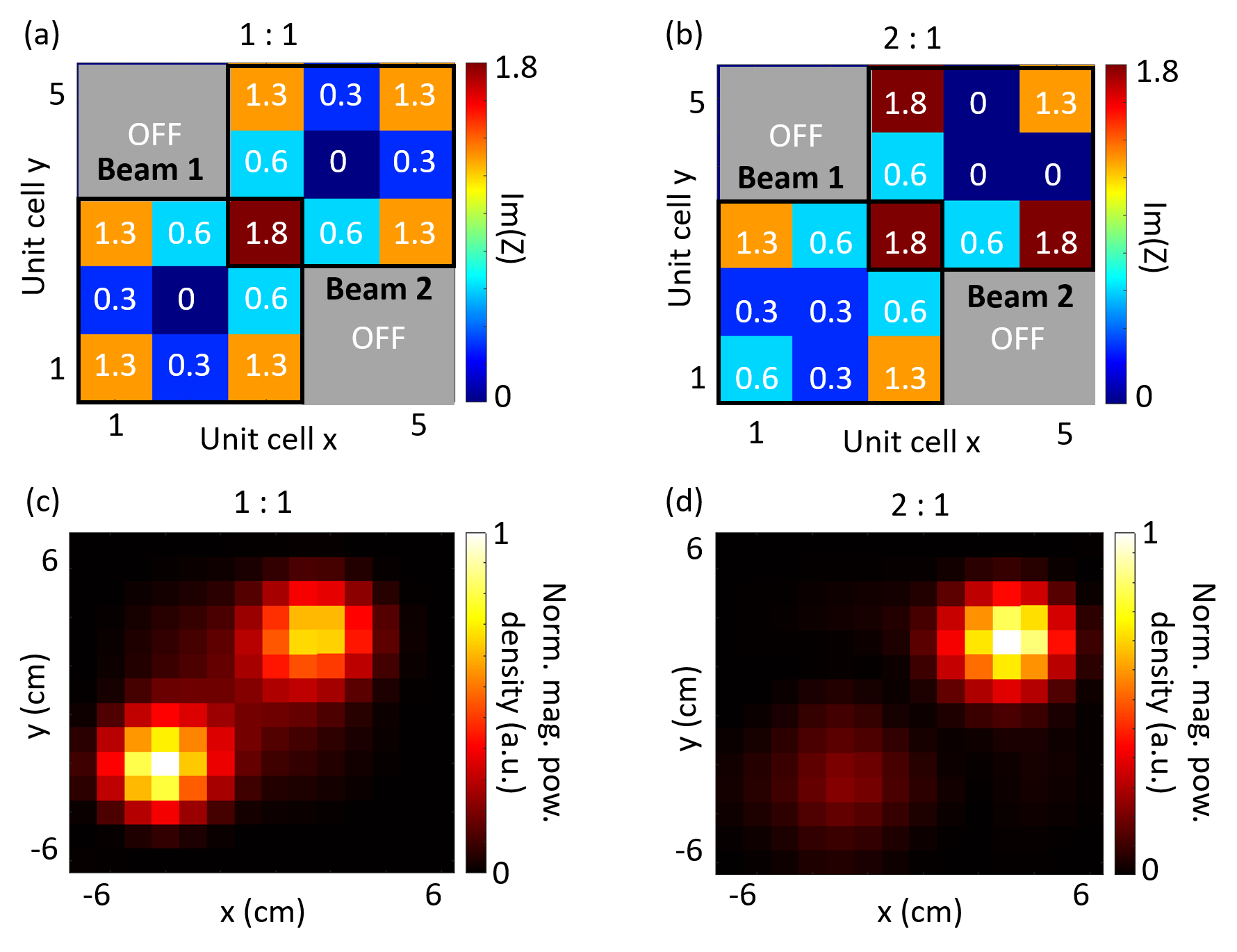}
	\caption{Two beams with tunable intensity ratios. Reactance distribution of (1) 1:1 and (2) 2:1. Magnetic power density of (1) 1:1 and (2) 2:1.}\label{FIG_22}
\end{figure}

\subsection{Multiple Rx coils with Tunable Power Division}
Multiple Rx coils placed at different locations may require different power. With the unique advantage of the metasurface, we can redistribute the magnetic flux into multiple beams with a controllable intensity ratio. We place two Rx coils at the positions of x = 3 cm, y = 3 cm, and x = -3 cm, y = -3 cm. We considered two scenarios: uniform Rx coils and non-uniform Rx coils. For the uniform case, we use two Rx coils both with a diameter of 4 cm. For the non-uniform case, we use two non-uniform Rx coils with diameters of 4 cm and 3.2 cm. Both the Rx coils are placed 3 cm from the metasurface in the z direction. All of the Rx coils are compensated to form resonance around 190 kHz.

The metasurface can tune the power division between multiple receivers while maintaining an enhanced or similar efficiency compared to the case without the metasurface. Ignoring the weak coupling between the Rx coils, the theoretical optimized targeted current distribution equals to the summation of the optimized targeted current distribution of the two Rx coils. Therefore, the equal power division case should be approximately the targeted current distribution for maximized overall efficiency of the two receivers. Due to their increased total reflected impedance to the Tx coil compared to the single receiver case, as shown in Table I, the measured overall efficiency of 71.9\% is higher than the single-user case of 63.5\% (Fig. 7). The slight difference in the power of the two receivers is because the two Rx coils have slightly different resonance frequencies. 

\begin{table}[!t]
	\renewcommand{\arraystretch}{1.3}
	\caption{Efficiency and power division for the uniform Rx case. The equal mode targets the two Rx coils with an intensity ratio of 1:1. The non-equal mode targets the two Rx coils with an intensity ratio of 2:1.}
	\centering
	\label{table_2}
	\resizebox{\columnwidth}{!}{
		\begin{tabular}{l l l l l}
			\hline\hline \\[-3mm]
			\multicolumn{1}{c}{} & \multicolumn{1}{c}{$\eta_1$} & \multicolumn{1}{c}{$\eta_2$} & \multicolumn{1}{c}{$\eta$} & \multicolumn{1}{c}{Power Ratio} \\[1.6ex] \hline
			    Without Metasurface  & 27.3\% & 28.6\% & 55.9\% & 1:1.05\\
                With (equal mode)  & 33.8\% & 38.1\% & 71.9\% & 1:1.13\\
                With (non-equal mode)  & 39.5\% & 10.9\% & 50.4\% & 1:0.28\\
                [1.4ex]
			\hline\hline
		\end{tabular}
  }
\end{table}

On the other hand, when the targeted beams have different intensities, the efficiency drop due to the broken spatial symmetry. The mode is not optimized for the maximum overall efficiency in this case. However, the system shows good controllability of the power divisions. Specifically, without the metasurface, the two receivers receive approximately the same power from the Tx, and the power ratio is 1:1.05. With the metasurface, in the case of the non-equal mode, different power is distributed to the two receivers. The two Rx coils receive 39.5\% and 10.9\% of the total transmitted power, respectively. The power ratio is changed to 1:0.28 with the metasurface. We can see that the overall efficiency in this case is no longer optimized but the receiver 1 received the highest power ratio among the three cases (Table I).

When the two Rx coils are non-uniform, the received power may be significantly different due to their non-balanced coupling to the Tx coil. The large Rx coil has a stronger reflected impedance to the Tx coil over the small Rx coil, and therefore, will gain the majority of the transmitted power. The small Rx coil can only obtain marginal power from the Tx coil due to the competing coupling from the large Rx coil. The metasurface can fix this unbalanced power division issue by enforcing a higher magnetic flux density to the small Rx coil. Without the metasurface, the small Rx coil only received maximumly 4.9\% of the total transmitted power, being 8.4 times smaller than the large Rx coil (41.1\% as shown in Fig. 11(a)). With the metasurface, the small Rx coil received 17.5\% of the total transmitted at 190 kHz, which is similar to the 14.6\% of the large Rx coil (Fig. 11(b)). The split peak of the large Rx coil's efficiency spectrum is because of the competing power division to the small Rx coil at the metasurface's operational frequency of around 190 kHz.

\begin{figure}[!t]\centering
	\includegraphics[width=9.5cm]{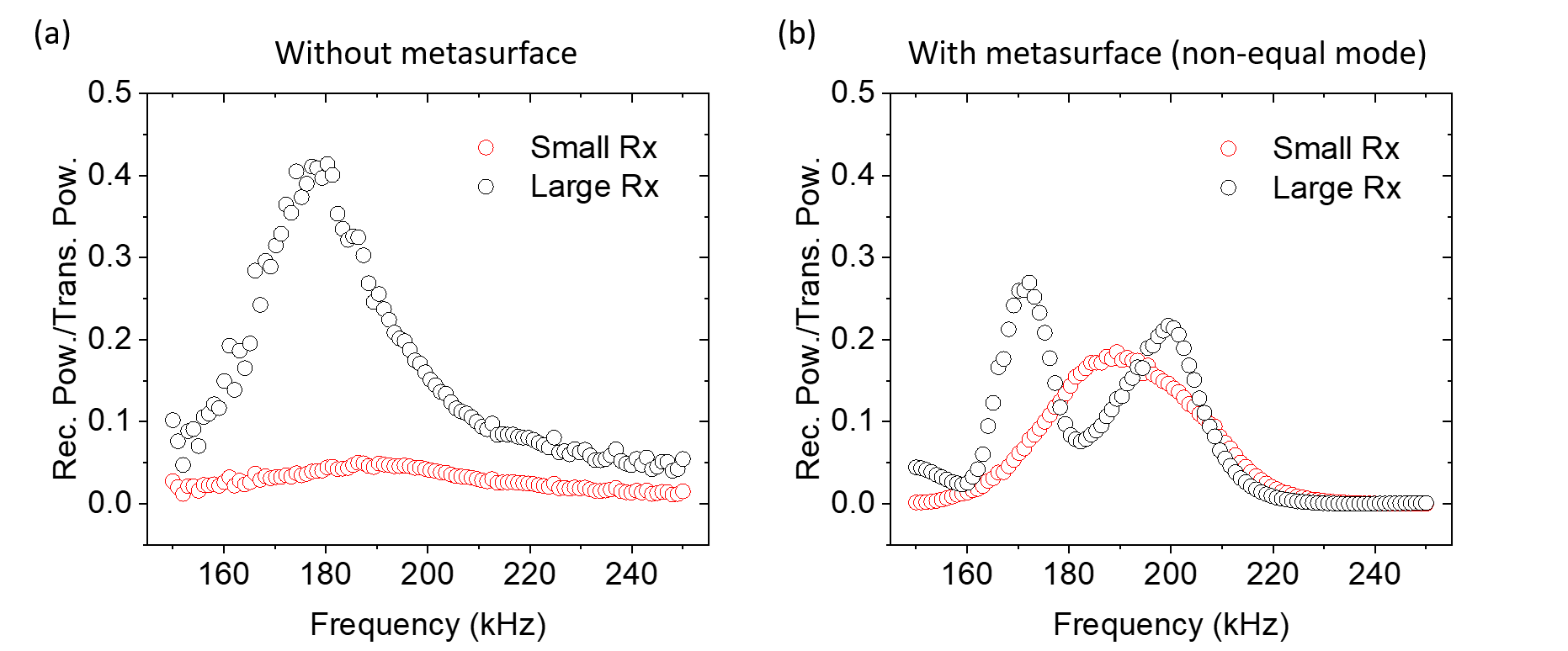}
	\caption{Received power ratio for the two receivers of different sizes. (a) The spectrum of the received power over the total transmitted power without metasurface. (b) The spectrum of the received power over the total transmitted power with metasurface. The metasurface is configured to the non-equal mode with a peak intensity ratio of 2:1.}\label{FIG_22}
\end{figure}

\section{Conclusion}

In this paper, we demonstrate a metasurface-enhanced wireless power transfer system that is compatible with Qi standard and supports multiple users with free-positioning capability. Our metasurface consists of a double-layer two-dimensional array of strongly coupled resonators that are capable of reshaping the magnetic power density to achieve free-positioning WPT. To guide our experiment, we provided a mutual inductance circuit model and inversely designed the metasurface's impedance configuration according to an optimized targeted current distribution. In the experiments, we have shown that our metasurface can achieve an efficiency enhancement of up to 4.6 times compared to the same condition without the metasurface. Without the metasurface, efficiency $>40\%$ only covers an area of around 5 by 5 cm; with the metasurface, we achieved a close-to-uniform efficiency $>70\%$, covering an area of around 10 by 10 cm. By taking the unique advantage of shaping the magnetic field with our metasurface, we demonstrated tuning the power division between the multiple Rx coils to compensate the receivers' size difference. Although the metasurface requires manual tuning, the principle can be easily extended to the automatically reconfigurable metasurfaces. For the multi-device case, the efficiency is sacrificed for non-unform power delivery as the targeted current distribution is not optimized. The optimized current distribution for non-unform power delivery can be studied in the future.

\section*{Acknowledgment}
The experiment in this paper is partially conducted in the Senior Design Laboratory of the Electrical and Computer Engineering department of the University of Illinois at Urbana-Champaign. Hanwei Wang is partially supported by Hong, MuCully, and Allen Fellowship, Yee Memorial Fund Fellowship, and Mavis Future Faculty Fellowship. 

\bibliographystyle{IEEEtranTIE}
\bibliography{Qi_Paper}\ 

\begin{thebibliography}{10}
\providecommand{\url}[1]{#1}
\csname url@samestyle\endcsname
\providecommand{\newblock}{\relax}
\providecommand{\bibinfo}[2]{#2}
\providecommand{\BIBentrySTDinterwordspacing}{\spaceskip=0pt\relax}
\providecommand{\BIBentryALTinterwordstretchfactor}{4}
\providecommand{\BIBentryALTinterwordspacing}{\spaceskip=\fontdimen2\font plus
\BIBentryALTinterwordstretchfactor\fontdimen3\font minus
  \fontdimen4\font\relax}
\providecommand{\BIBforeignlanguage}[2]{{%
\expandafter\ifx\csname l@#1\endcsname\relax
\typeout{** WARNING: IEEEtran.bst: No hyphenation pattern has been}%
\typeout{** loaded for the language `#1'. Using the pattern for}%
\typeout{** the default language instead.}%
\else
\language=\csname l@#1\endcsname
\fi
#2}}
\providecommand{\BIBdecl}{\relax}
\BIBdecl

\bibitem{hui2013planar}
S.~Hui, ``Planar wireless charging technology for portable electronic products
  and qi,'' \emph{Proc. IEEE}, vol. 101, no.~6, pp. 1290--1301, 2013.

\bibitem{li2014wireless}
S.~Li and C.~C. Mi, ``Wireless power transfer for electric vehicle
  applications,'' \emph{IEEE journal of emerging and selected topics in power
  electronics}, vol.~3, no.~1, pp. 4--17, 2014.

\bibitem{chen2021stability}
K.~Chen, K.~W.~E. Cheng, Y.~Yang, and J.~Pan, ``Stability improvement of
  dynamic ev wireless charging system with receiver-side control considering
  coupling disturbance,'' \emph{Electronics}, vol.~10, no.~14, p. 1639, 2021.

\bibitem{chen2016cost}
W.~Chen, C.~Liu, C.~H. Lee, and Z.~Shan, ``Cost-effectiveness comparison of
  coupler designs of wireless power transfer for electric vehicle dynamic
  charging,'' \emph{Energies}, vol.~9, no.~11, p. 906, 2016.

\bibitem{sample2007design}
A.~P. Sample, D.~J. Yeager, P.~S. Powledge, and J.~R. Smith, ``Design of a
  passively-powered, programmable sensing platform for uhf rfid systems,'' in
  \emph{2007 IEEE international Conference on RFID}, pp. 149--156.\hskip 1em
  plus 0.5em minus 0.4em\relax IEEE, 2007.

\bibitem{ho2014wireless}
J.~S. Ho, A.~J. Yeh, E.~Neofytou, S.~Kim, Y.~Tanabe, B.~Patlolla, R.~E. Beygui,
  and A.~S. Poon, ``Wireless power transfer to deep-tissue microimplants,''
  \emph{Proceedings of the National Academy of Sciences}, vol. 111, no.~22, pp.
  7974--7979, 2014.

\bibitem{ozaki2021wireless}
T.~Ozaki, N.~Ohta, T.~Jimbo, and K.~Hamaguchi, ``A wireless
  radiofrequency-powered insect-scale flapping-wing aerial vehicle,''
  \emph{Nature Electronics}, vol.~4, no.~11, pp. 845--852, 2021.

\bibitem{kiziroglou2017acoustic}
M.~Kiziroglou, D.~Boyle, S.~Wright, and E.~Yeatman, ``Acoustic power delivery
  to pipeline monitoring wireless sensors,'' \emph{Ultrasonics}, vol.~77, pp.
  54--60, 2017.

\bibitem{liu2016charging}
Q.~Liu, J.~Wu, P.~Xia, S.~Zhao, W.~Chen, Y.~Yang, and L.~Hanzo, ``Charging
  unplugged: Will distributed laser charging for mobile wireless power transfer
  work?'' \emph{IEEE Vehicular Technology Magazine}, vol.~11, no.~4, pp.
  36--45, 2016.

\bibitem{zhang2016efficiency}
C.~Zhang, D.~Lin, and S.~R. Hu, ``Efficiency optimization method of inductive
  coupling wireless power transfer system with multiple transmitters and single
  receiver,'' in \emph{2016 IEEE Energy Conversion Congress and Exposition
  (ECCE)}, pp. 1--6.\hskip 1em plus 0.5em minus 0.4em\relax IEEE, 2016.

\bibitem{hui2013critical}
S.~Y.~R. Hui, W.~Zhong, and C.~K. Lee, ``A critical review of recent progress
  in mid-range wireless power transfer,'' \emph{IEEE Transactions on Power
  Electronics}, vol.~29, no.~9, pp. 4500--4511, 2013.

\bibitem{karalis2008efficient}
A.~Karalis, J.~D. Joannopoulos, and M.~Solja{\v{c}}i{\'c}, ``Efficient wireless
  non-radiative mid-range energy transfer,'' \emph{Annals of physics}, vol.
  323, no.~1, pp. 34--48, 2008.

\bibitem{assawaworrarit2017robust}
S.~Assawaworrarit, X.~Yu, and S.~Fan, ``Robust wireless power transfer using a
  nonlinear parity--time-symmetric circuit,'' \emph{Nature}, vol. 546, no.
  7658, pp. 387--390, 2017.

\bibitem{hui2005new}
S.~R. Hui and W.~W. Ho, ``A new generation of universal contactless battery
  charging platform for portable consumer electronic equipment,'' \emph{IEEE
  Transactions on Power Electronics}, vol.~20, no.~3, pp. 620--627, 2005.

\bibitem{jadidian2014magnetic}
J.~Jadidian and D.~Katabi, ``Magnetic mimo: How to charge your phone in your
  pocket,'' in \emph{Proceedings of the 20th annual international conference on
  Mobile computing and networking}, pp. 495--506, 2014.

\bibitem{wang2021comparative}
H.~Wang, C.~Zhang, Y.~Yang, H.~W.~R. Liang, and S.~Y.~R. Hui, ``A comparative
  study on overall efficiency of two-dimensional wireless power transfer
  systems using rotational and directional methods,'' \emph{IEEE Transactions
  on Industrial Electronics}, vol.~69, no.~1, pp. 260--269, 2021.

\bibitem{wang2020robust}
N.~X. Wang, H.-W. Wang, J.~Mei, S.~Mohammadi, J.~Moon, J.~H. Lang, and J.~L.
  Kirtley, ``Robust 3-d wireless power transfer system based on rotating fields
  for multi-user charging,'' \emph{IEEE Transactions on Energy Conversion},
  vol.~36, no.~2, pp. 693--702, 2020.

\bibitem{shi2015wireless}
L.~Shi, Z.~Kabelac, D.~Katabi, and D.~Perreault, ``Wireless power hotspot that
  charges all of your devices,'' in \emph{Proceedings of the 21st Annual
  International Conference on Mobile Computing and Networking}, pp. 2--13,
  2015.

\bibitem{waters2015power}
B.~H. Waters, B.~J. Mahoney, V.~Ranganathan, and J.~R. Smith, ``Power delivery
  and leakage field control using an adaptive phased array wireless power
  system,'' \emph{IEEE Transactions on Power Electronics}, vol.~30, no.~11, pp.
  6298--6309, 2015.

\bibitem{graham2019wireless}
C.~S. Graham, D.~B. Karanikos, D.~R. Kasar, P.~J. Thompson, E.~S. Jol, G.~S.
  Haug, K.~R.~F. Larsson, and S.~A. Kowalski, ``Wireless charging mats for
  portable electronic devices,'' Apr.~30 2019, uS Patent 10,277,043.

\bibitem{holloway2012overview}
C.~L. Holloway, E.~F. Kuester, J.~A. Gordon, J.~O'Hara, J.~Booth, and D.~R.
  Smith, ``An overview of the theory and applications of metasurfaces: The
  two-dimensional equivalents of metamaterials,'' \emph{IEEE antennas and
  propagation magazine}, vol.~54, no.~2, pp. 10--35, 2012.

\bibitem{wang2021demand}
H.~Wang, H.-K. Huang, Y.-S. Chen, and Y.~Zhao, ``On-demand field shaping for
  enhanced magnetic resonance imaging using an ultrathin reconfigurable
  metasurface,'' \emph{View}, vol.~2, no.~3, p. 20200099, 2021.

\bibitem{zhao2022ultrathin}
Y.~Zhao, H.~Wang, and Y.-S. Chen, ``Ultrathin reconfigurable metamaterial for
  signal enhancement of magnetic resonance imaging,'' Jun.~30 2022, uS Patent
  App. 17/560,505.

\bibitem{wang2021wearable}
H.~Wang, Y.-S. Chen, and Y.~Zhao, ``A wearable metasurface for high efficiency,
  free-positioning omnidirectional wireless power transfer,'' \emph{New Journal
  of Physics}, vol.~23, no.~12, p. 125003, 2021.

\bibitem{ranaweera2019active}
A.~Ranaweera, T.~S. Pham, H.~N. Bui, V.~Ngo, and J.-W. Lee, ``An active
  metasurface for field-localizing wireless power transfer using dynamically
  reconfigurable cavities,'' \emph{Scientific reports}, vol.~9, no.~1, p.
  11735, 2019.

\bibitem{shi2016large}
X.~Shi and J.~R. Smith, ``Large area wireless power via a planar array of
  coupled resonators,'' in \emph{2016 International Workshop on Antenna
  Technology (iWAT)}, pp. 200--203.\hskip 1em plus 0.5em minus 0.4em\relax
  IEEE, 2016.

\end{thebibliography}
\vspace{-1.5cm}
\begin{IEEEbiography}[{\includegraphics[width=1in,height=1.25in,clip,keepaspectratio]{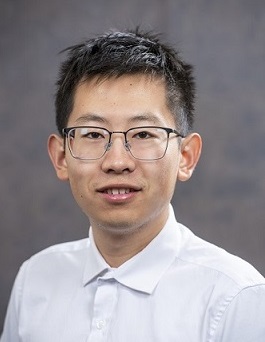}}]
{Hanwei Wang} received his B.S. degree in physics from the Department of Physics, Tsinghua University, Beijing, China, in 2019. He is currently working toward the Ph.D. degree in electrical engineering with the Department of Electrical and Computer Engineering, University of Illinois at Urbana-Champaign, Champaign, IL, USA. His research interests include metamaterials, optical force microscopy, magnetic resonance imaging, and wireless power transfer.
\end{IEEEbiography}
\vspace{-1cm}
\begin{IEEEbiography}[{\includegraphics[width=1in,clip,keepaspectratio]{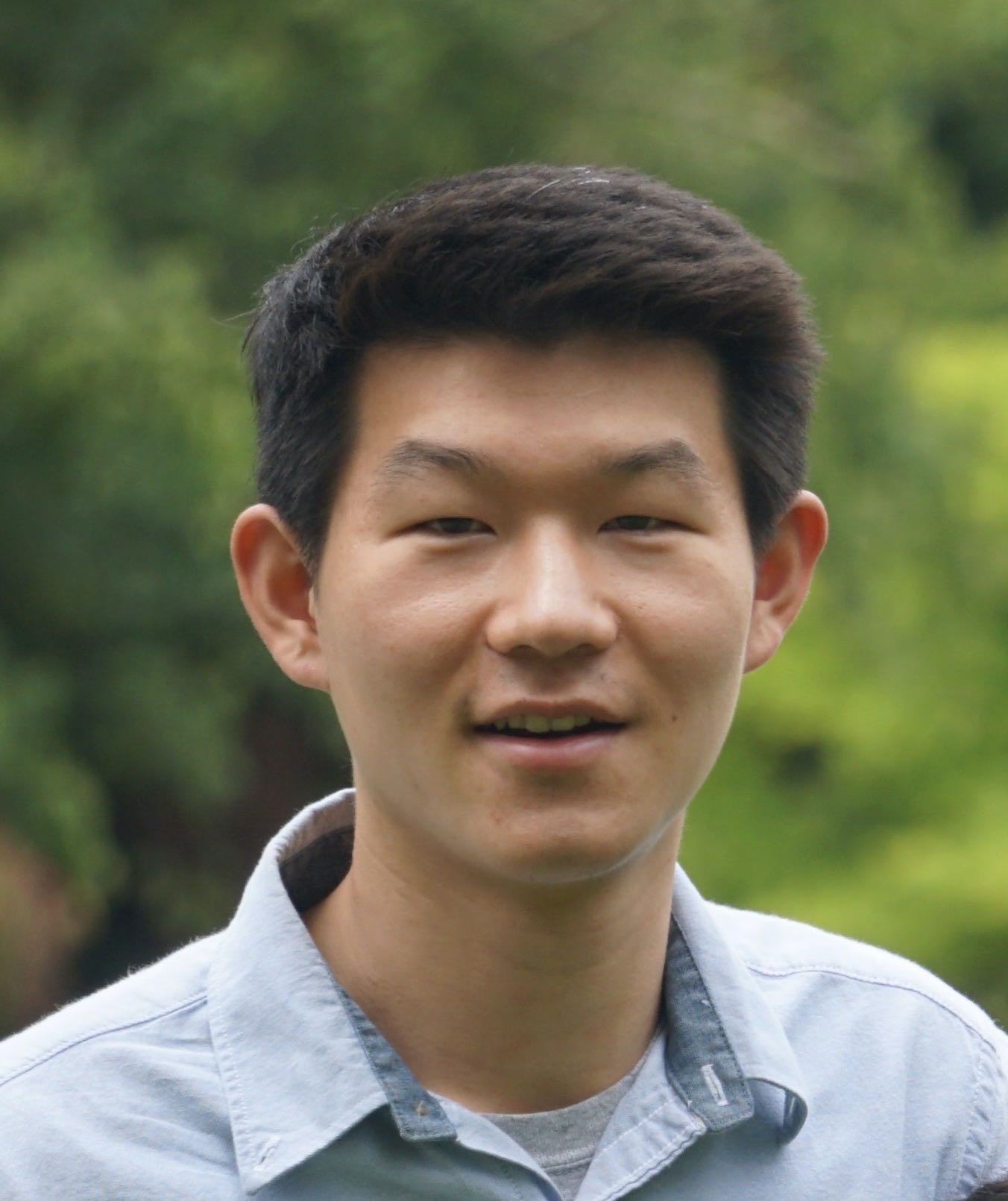}}]
{Joshua Yu} is an undergraduate student in the Department of Electrical and Computer Engineering, University of Illinois at Urbana-Champaign, Champaign, IL, USA. His research interests include wireless power transfer, metamaterials, and antennas.
\end{IEEEbiography}

\begin{IEEEbiography}[{\includegraphics[width=1in,clip,keepaspectratio]{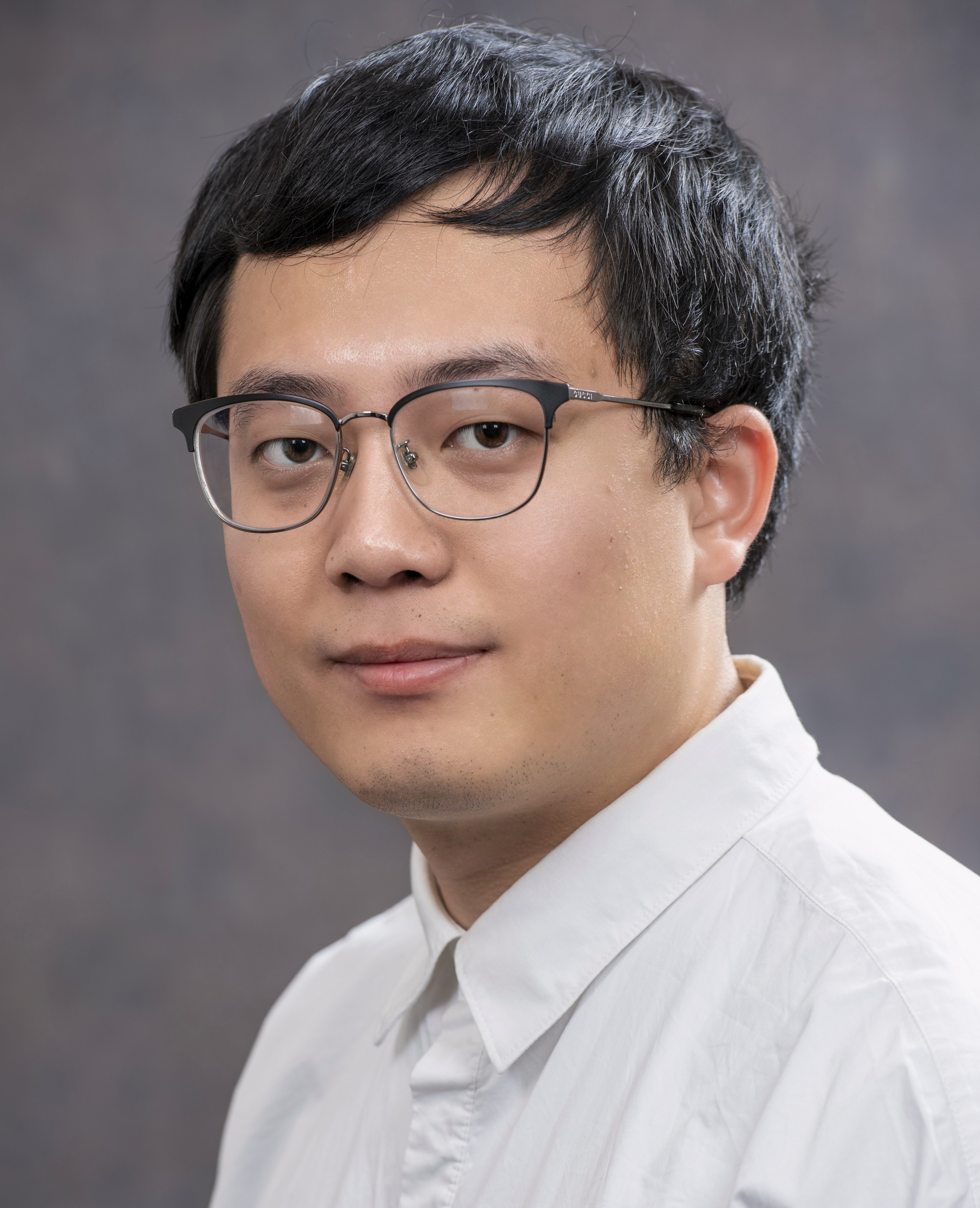}}]
{Xiaodong Ye} (S'23) was born in Ningbo, China. He received his B.Eng. degree in Electrical and Electronics in 2020 from University of Liverpool, Liverpool, UK, and M.Eng. degree in Electrical and Computer Engineering in 2022 from University of Illinois Urbana-Champaign, Urbana, USA, where he is currently working toward the Ph.D. degree in Electrical and Computer Engineering. His current research interests include metamaterial-enhanced wireless power transfer and near-field magnetic interaction.
\end{IEEEbiography}

\begin{IEEEbiography}[{\includegraphics[width=1in,clip,keepaspectratio]{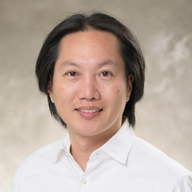}}]
{Yun-Sheng Chen} is an assistant professor at the University of Illinois, Urbana-Champaign, Department of Electrical and Computer Engineering. He is affiliated with the Beckman Institute and has a courtesy appointment with the Department of Bioengineering, and the Department of Biomedical and Translational Sciences at Carle Illinois College of Medicine. Prof. Chen's lab works on medical imaging augmented with machine learning and nanotechnology. His research interests include multimodality imaging, radio-frequency acoustic imaging, photoacoustic imaging, ultrasound imaging, and molecular imaging. 
\end{IEEEbiography}

\begin{IEEEbiography}[{\includegraphics[width=1in,clip,keepaspectratio]{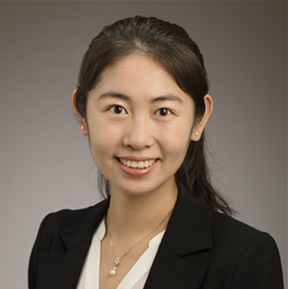}}]
{Yang Zhao} is an assistant professor at the University of Illinois, Urbana-Champaign, Department of Electrical and Computer Engineering. She is affiliated with the Micro and Nanotechnology Laboratory and has a courtesy appointment with the Department of Bioengineering, and the Institute for Genomic Biology at UIUC. Prof. Zhao directs the BioNanophotonics lab at UIUC. Her research group is interested in studying light-matter interactions across subwavelength to wavelength scales. Her lab develops metamaterials and nanophotonic devices to manipulate the near-field optical forces and electromagnetic field distribution with applications in imaging, biosensing, actuation, and wireless power transfer. 
\end{IEEEbiography}

\end{document}